\documentclass{llncs}
\usepackage{latexsym}
\usepackage{graphicx}
\newcommand{\be}{\begin{equation}}
\newcommand{\ee}{\end{equation}}
\newcommand{\bea}{\begin{eqnarray}}
\newcommand{\eea}{\end{eqnarray}}

\newcommand{\bee}{\begin{eqnarray*}}
\newcommand{\eee}{\end{eqnarray*}}
\title{\bf Synthesis of all Maximum Length Cellular Automata of Cell Size up to $12$}
\author{Jaydeb Bhaumik \\}
\institute{\ Haldia Institute of Technology, Haldia-$721657$,
India\\}
\date{}
\begin{document}
\maketitle
\begin{abstract}
Maximum length CA has wide range of applications in design of
linear block code, cryptographic primitives and VLSI testing
particularly in Built-In-Self-Test. In this paper, an algorithm to
compute all $n$-cell maximum length CA-rule vectors is proposed.
Also rule vectors for each primitive polynomial in $GF(2^2)$ to
$GF(2^{12})$ have been computed by simulation and they have been
listed.Programmable rule vectors based maximum length
CA can be used to design cryptographic primitives.\\

\end{abstract}

{\section{Introduction}\setcounter{equation}{0}} \noindent
 A Cellular Automata (CA) consist of a number of cells arranged in a
regular manner. Each cell consists of a storage element (D
flip-flop) and a combinational logic implementing the next-state
function. CA is universally accepted as a very good generator of
pseudo random sequences. It is also very well suited for VLSI
design due to its regular structure.If the combinational logic of
a CA cell only involves XOR logic, then it is called a linear CA.
For a three neighborhood one dimensional CA, the combinational
logic implementing the next state is $s_i (t+1) =  f(s_{i-1}(t),
s_i(t), s_{i+1}(t))$. Where $s_i(t)$ is the output state of the
$i_{th}$ cell at $t_{th}$ time step. $s_{i-1}(t)$ and $s_{i+1}(t)$
are the output states of left and right neighbors of $i_{th}$ cell
and $f$ denotes the local transition function realized with a
combinational logic and is known as a rule of the CA. A CA is said
to be hybrid if the rules of different cells vary. An $n$-cell
maximum length CA is characterized by the presence of a cycle of
length $2^n -1$ with all non-zero states. In case of a maximum
length CA, it has a characteristic polynomial which is primitive.
CA-rules $90$ and $150$ have been considered. The combinational
logic for rule $90$
and rule $150$ are as follows.\\
Rule $90$ :    $s_i (t+1) =  s_{i-1}(t)\oplus s_{i+1} (t) $\\
Rule $150$ :   $s_i (t+1) =  s_{i -1}(t)\oplus s_i(t)\oplus s_{i+1}(t)$\\
where $s_i(t)$ is the output state of the $i$-th cell at time $t$.

Efficient characterization of $1D$ CA based on matrix algebra and
its application in error correcting codes, cryptography
\cite{kar94}and VLSI testing is available in \cite{chaudhuri97}.
The characteristic matrix of a linear CA operating over $GF(2)$ is
a matrix that describes the behavior of the CA. We can calculate
the next state of the CA by multiplying the characteristic matrix
by the present state of the CA. A characteristic matrix is
constructed as: $T[i, j] = 1$, if the next state of the $i_{th}$
cell depends on the $j_{th}$ cell and $T[i, j] = 0$, otherwise.

Only one rule vector for each $n$-length CA has been provided in
\cite{chaudhuri97}. A new architectural design of CA-based codec
based on linear maximum length CA has been proposed in
\cite{bhaumik10}. In \cite{cattell95} authors proposed an
algorithm for determining minimal cost $n$-cell maximum length CA
of degree up to $500$. Programmable rule vectors based linear
maximum length CA has many applications in the design of
cryptographic primitives. In \cite{bhaumik09} one such application
has been mentioned, where programmable linear maximum length CA
has been used to design an integrated scheme for both error
correction and message authentication. Therefore, designer needs
list of maximum length CA-rule vectors for a particular cell size.

\section{Method and Result} The algorithm of determining whether a
given $n$- cell CA has a maximum length cycle is as follows.
\begin{enumerate}
\item Take $n\times n$ tridiagonal matrix with all non-zero elements are
$1$
\item Change main diagonal sequentially by one of the
$2^n$ combinations
\item Compute the characteristic polynomial corresponding to the
$n\times n$ constructed matrix
\item Calculate the number of non-zero coefficient in the characteristic polynomial
and if number of coefficients is even then go to step $2$
\item Check the coefficients of $x^n$ and $x^0$, if they are zeros
then go to step $2$.
\item Check if the characteristic polynomial matches with any one of the list  of
primitive polynomials.
\item If matches then corresponding main diagonal of the matrix represents the
maximum length CA-rule vector
\end{enumerate}

In Table under the caption 'CA-rule vector', $`0'$ and $`1'$
correspond to rule $90$ and $150$ respectively. Under caption
'Primitive poly.' the entries represent primitive polynomial in
binary format. It has been observed that mirror image of each rule
vector corresponds to same primitive polynomial. For example in
$8$-cell CA, $00000110$ and $01100000$ are two rule vectors for
primitive polynomial $x^8 + x^4 + x^3 + x^2 + 1$ $(100011101)$,
where rule vectors are mirror image of each other

\begin{tabular}{|c|c|c|}
\hline
\# cells & Primitive Poly. & CA-rule vector\\
\hline $2$ & $111$ & $10$ \\
\hline $3$ & $1011$ & $110$\\
\cline{2-3} $$ & $1101$ & $100$\\
\hline $4$ & $10011$ & $1010$\\
\cline{2-3} $$ & $11001$ & $1101$\\
\hline  $5$ & $100101$ & $11100$\\
\cline{2-3} $$ & $101001$ & $10000$\\
\cline{2-3} $$ & $101111$ & $01100$\\
\cline{2-3} $$ & $110111$ & $10011$\\
\cline{2-3} $$ & $111011$ & $11000$\\
\cline{2-3} $$ & $111101$ & $11110$\\
\hline  $6$ & $1000011$ & $000110$\\
\cline{2-3} $$ & $1011011$ & $101110$\\
\cline{2-3} $$ & $1100001$ & $011010$\\
\cline{2-3} $$ & $1100111$ & $100101$\\
\cline{2-3} $$ & $1101101$ & $101010$\\
\cline{2-3} $$ & $1110011$ & $100000$\\
\hline $7$ & $10000011$ & $1011001$\\
\cline{2-3} $$ & $ 10001001$ & $0111010$\\
\cline{2-3}$$ & $10001111$ & $1110001$\\
\cline{2-3} $$ & $10010001$ & $1110100$\\
\cline{2-3} $$ & $10011101$ & $1101010$\\
\cline{2-3} $$ & $10100111$ & $0010010$\\
\cline{2-3} $$ & $10101011$ & $1101111$\\
\cline{2-3} $$ & $10111001$ & $1001000$\\
\cline{2-3} $$ & $10111111$ & $1000010$\\
\cline{2-3} $$ & $11000001$ & $0010000$\\
\cline{2-3} $$ & $11001011$ & $1011011$\\
\cline{2-3} $$ & $11010011$ & $0110111$\\
\cline{2-3} $$ & $11010101$ & $1011110$\\
\cline{2-3} $$ & $11100101$ & $1010100$\\
\cline{2-3} $$ & $11101111$ & $1101000$\\
\cline{2-3} $$ & $11110001$ & $1000101$\\
\cline{2-3} $$ & $11110111$ & $0001110$\\
\cline{2-3} $$ & $11111101$ & $0100110$\\
\hline
\end{tabular}
\quad
\begin{tabular}{|c|c|c|}
\hline
\# cells & Primitive Poly. & CA-rule vector\\
\hline     $8$ & $100011101$ & $00000110$\\
\cline{2-3} $$ & $ 100101011$ & $01011111$\\
\cline{2-3} $$ & $100101101$ & $01110111$\\
\cline{2-3} $$ & $101001101$ & $01101100$\\
\cline{2-3} $$ & $101011111$ & $10010011$\\
\cline{2-3} $$ & $101100011$ & $11010010$\\
\cline{2-3} $$ & $101100101$ & $00101101$\\
\cline{2-3} $$ & $101101001$ & $00001111$\\
\cline{2-3} $$ & $101110001$ & $00111001$\\
\cline{2-3} $$ & $110000111$ & $11101111$\\
\cline{2-3} $$ & $110001101$ & $00101010$\\
\cline{2-3} $$ & $110101001$ & $01000101$\\
\cline{2-3} $$ & $111000011$ & $01011101$\\
\cline{2-3} $$ & $111001111$ & $01011011$\\
\cline{2-3} $$ & $111100111$ & $11001011$\\
\cline{2-3} $$ & $111110101$ & $11010101$\\
\hline $9$ & $1000010001$  & $001110010$\\
\cline{2-3} $$ & $1000011011$  &  $000101110$\\
\cline{2-3} $$ & $1000100001$  &  $100110001$\\
\cline{2-3} $$ & $1000101101$  &  $100001101$\\
\cline{2-3} $$ & $1000110011$  &  $001101010$\\
\cline{2-3} $$ & $1001011001$  &  $010100011$\\
\cline{2-3} $$ & $1001011111$  &  $001100101$\\
\cline{2-3} $$ & $1001101001$  &  $011100001$\\
\cline{2-3} $$ & $1001101111$  &  $000100111$\\
\cline{2-3} $$ & $1001110111$  &  $011111111$\\
\cline{2-3} $$ & $1001111101$  &  $000011101$\\
\cline{2-3} $$ & $1010000111$  &  $010111110$\\
\cline{2-3} $$ & $1010010101$  &  $000000110$\\
\cline{2-3} $$ & $1010100011$  &  $110110011$\\
\cline{2-3} $$ & $1010100101$  &  $101111001$\\
\cline{2-3} $$ & $1010101111$  &  $001001000$\\
\cline{2-3} $$ & $1010110111$  &  $101100111$\\
\cline{2-3} $$ & $1010111101$  &  $101001111$\\
\cline{2-3} $$ & $1011001111$  &  $011010111$\\
\hline
\end{tabular}

\begin{tabular}{|c|c|c|}
\hline
\# cells & Primitive Poly. & CA-rule vector\\
\hline $9$ & $1011010001$  &  $010000001$\\
\cline{2-3} $$ & $1011011011$  &  $101011110$\\
\cline{2-3} $$ & $1011110101$  &  $001111011$\\
\cline{2-3} $$ & $1011111001$  &  $001011111$\\
\cline{2-3} $$ & $1100010011$  &  $101100011$\\
\cline{2-3} $$ & $1100010101$  &  $100011011$\\
\cline{2-3} $$ & $1100011111$  &  $100010111$\\
\cline{2-3} $$ & $1100100011$  &  $110010101$\\
\cline{2-3} $$ & $1100110001$  &  $011100110$\\
\cline{2-3} $$ & $1100111011$  &  $010011110$\\
\cline{2-3} $$ & $1101001111$  &  $010110011$\\
\cline{2-3} $$ & $1101011011$  &  $011001101$\\
\cline{2-3} $$ & $1101100001$  &  $000101111$\\
\cline{2-3} $$ & $1101101011$  &  $011110001$\\
\cline{2-3} $$ & $1101101101$  &  $000110111$\\
\cline{2-3} $$ & $1101110011$  &  $000000001$\\
\cline{2-3} $$ & $1101111111$  &  $010001111$\\
\cline{2-3} $$ & $1110000101$  &  $000001011$\\
\cline{2-3} $$ & $1110001111$  &  $001000011$\\
\cline{2-3} $$ & $1110110101$  &  $010100001$\\
\cline{2-3} $$ & $1110111001$  &  $011111101$\\
\cline{2-3} $$ & $1111000111$  &  $111011011$\\
\cline{2-3} $$ & $1111001011$  &  $010000110$\\
\cline{2-3} $$ & $1111001101$  &  $001001100$\\
\cline{2-3} $$ & $1111010101$  &  $000011010$\\
\cline{2-3} $$ & $1111011001$  &  $000110010$\\
\cline{2-3} $$ & $1111100011$  &  $000001110$\\
\cline{2-3} $$ & $1111101001$  &  $100111111$\\
\cline{2-3} $$ & $1111111011$  &  $101000001$\\
\hline $10$ & $10000001001$  & $1100001111$\\
\cline{2-3} $$ & $10000011011$  &  $1001110101$\\
\cline{2-3} $$ & $10000100111$  &  $1010100111$\\
\cline{2-3} $$ & $10000101101$  &  $1001010111$\\
\cline{2-3} $$ & $10001100101$  &  $1000111011$\\
\hline
\end{tabular}
\quad
\begin{tabular}{|c|c|c|}
\hline
\# cells & Primitive Poly. & CA-rule vector\\
\hline $10$ & $10001101111$  &  $0001000010$\\
\cline{2-3} $$ & $10010000001$  &  $0000111111$\\
\cline{2-3} $$ & $10010001011$  &  $0111011001$\\
\cline{2-3} $$ & $10011000101$  &  $0001111011$\\
\cline{2-3} $$ & $10011010111$  &  $0011001111$\\
\cline{2-3} $$ & $10011100111$  &  $0111100011$\\
\cline{2-3} $$ & $10011110011$  &  $0100011111$\\
\cline{2-3} $$ & $10011111111$  &  $0011111001$\\
\cline{2-3} $$ & $10100001101$  &  $1011000001$\\
\cline{2-3} $$ & $10100011001$  &  $0101000110$\\
\cline{2-3} $$ & $10100100011$  &  $0010100110$\\
\cline{2-3} $$ & $10100110001$  &  $0001101010$\\
\cline{2-3} $$ & $10100111101$  &  $0001010110$\\
\cline{2-3} $$ & $10101000011$  &  $0100101010$\\
\cline{2-3} $$ & $10101010111$  &  $1111010111$\\
\cline{2-3} $$ & $10101101011$  &  $1110111101$\\
\cline{2-3} $$ & $10110000101$  &  $0110000011$\\
\cline{2-3} $$ & $10110001111$  &  $0000011101$\\
\cline{2-3} $$ & $10110010111$  &  $0001110001$\\
\cline{2-3} $$ & $10110100001$  &  $1001010010$\\
\cline{2-3} $$ & $10111000111$  &  $0010011001$\\
\cline{2-3} $$ & $10111100101$  &  $1000010110$\\
\cline{2-3} $$ & $10111110111$  &  $1000100110$\\
\cline{2-3} $$ & $10111111011$  &  $0000001111$\\
\cline{2-3} $$ & $11000010011$  &  $1100101111$\\
\cline{2-3} $$ & $11000010101$  &  $0000101100$\\
\cline{2-3} $$ & $11000100101$  &  $0001100010$\\
\cline{2-3} $$ & $11000110111$  &  $1100000001$\\
\cline{2-3} $$ & $11001000011$  &  $0001001010$\\
\cline{2-3} $$ & $11001001111$  &  $0010001010$\\
\cline{2-3} $$ & $11001011011$  &  $1101110101$\\
\cline{2-3} $$ & $11001111001$  &  $1011011101$\\
\cline{2-3} $$ & $11001111111$  &  $0100010010$\\
\cline{2-3} $$ & $11010001001$  &  $0110111011$\\
\hline
\end{tabular}

\begin{tabular}{|c|c|c|}
\hline
\# cells & Primitive Poly. & CA-rule vector\\
\hline  $10$ & $11010110101$  &  $1011101110$\\
\cline{2-3} $$ & $11011000001$  &  $1000001010$\\
\cline{2-3} $$ & $11011010011$  &  $0010111111$\\
\cline{2-3} $$ & $11011011111$  &  $0100010001$\\
\cline{2-3} $$ & $11011111101$  &  $0111110011$\\
\cline{2-3} $$ & $11100010111$  &  $0011000111$\\
\cline{2-3} $$ & $11100011101$  &  $0011100011$\\
\cline{2-3} $$ & $11100100001$  &  $1101001010$\\
\cline{2-3} $$ & $11100111001$  &  $0110000111$\\
\cline{2-3} $$ & $11101000111$  &  $1100100110$\\
\cline{2-3} $$ & $11101001101$  &  $0001101101$\\
\cline{2-3} $$ & $11101010101$  &  $0011011001$\\
\cline{2-3} $$ & $11101011001$  &  $0100100111$\\
\cline{2-3} $$ & $11101100011$  &  $0101010101$\\
\cline{2-3} $$ & $11101111101$  &  $0110110001$\\
\cline{2-3} $$ & $11110001101$  &  $1100000111$\\
\cline{2-3} $$ & $11110010011$  &  $0101010110$\\
\cline{2-3} $$ & $11110110001$  &  $0011100110$\\
\cline{2-3} $$ & $11111011011$  &  $1100011001$\\
\cline{2-3} $$ & $11111110011$  &  $0001111100$\\
\cline{2-3} $$ & $11111111001$  &  $0110010110$\\
\hline $11$ & $100000000101$ &  $01000011010$    \\
\cline{2-3}$$ & $100000010111$ &  $11110101011$  \\
\cline{2-3}$$ & $100000101011$ &  $01000110010$  \\
\cline{2-3}$$ & $100000101101$ &  $01101111110$  \\
\cline{2-3}$$ & $100001000111$ &  $00110010010$  \\
\cline{2-3}$$ & $100001100011$ &  $10001000011$  \\
\cline{2-3}$$ & $100001100101$ &  $00110100010$  \\
\cline{2-3}$$ & $100001110001$ &  $00010110010$  \\
\cline{2-3}$$ & $100001111011$ &  $00100010110$  \\
\cline{2-3}$$ & $100010001101$ &  $11101001111$  \\
\cline{2-3}$$ & $100010010101$ &  $00110011000$  \\
\cline{2-3}$$ & $100010011111$ &  $10100001001$  \\
\cline{2-3}$$ & $100010101001$ &  $11100110111$  \\
\cline{2-3}$$ & $100010110001$ &  $00011100100$  \\
\hline
\end{tabular}
\quad
\begin{tabular}{|c|c|c|}
\hline
\# cells & Primitive Poly. & CA-rule vector\\
\hline $11$ & $100011001111$ &  $00001001110$  \\
\cline{2-3}$$ & $100011010001$ &  $00001110010$  \\
\cline{2-3}$$ & $100011100001$ &  $00100101010$  \\
\cline{2-3}$$ & $100011100111$ &  $01001010100$  \\
\cline{2-3}$$ & $100011101011$ &  $00001100110$  \\
\cline{2-3}$$ & $100011110101$ &  $10011110111$  \\
\cline{2-3}$$ & $100100001101$ &  $01111011101$  \\
\cline{2-3}$$ & $100100010011$ &  $00000110011$  \\
\cline{2-3}$$ & $100100100101$ &  $01000101001$  \\
\cline{2-3}$$ & $100100101001$ &  $00000100111$  \\
\cline{2-3}$$ & $100100111011$ &  $01001000101$  \\
\cline{2-3}$$ & $100100111101$ &  $00100000111$  \\
\cline{2-3}$$ & $100101000101$ &  $01101101111$  \\
\cline{2-3}$$ & $100101001001$ &  $01101110111$  \\
\cline{2-3}$$ & $100101010001$ &  $01000000111$  \\
\cline{2-3}$$ & $100101011011$ &  $00001110001$  \\
\cline{2-3}$$ & $100101110011$ &  $11000000110$  \\
\cline{2-3}$$ & $100101110101$ &  $01011101111$  \\
\cline{2-3}$$ & $100101111111$ &  $00010101001$  \\
\cline{2-3}$$ & $100110000011$ &  $00010001011$  \\
\cline{2-3}$$ & $100110001111$ &  $10000010110$  \\
\cline{2-3}$$ & $100110101011$ &  $01100010001$  \\
\cline{2-3}$$ & $100110101101$ &  $00110111111$  \\
\cline{2-3}$$ & $100110111001$ &  $01001100001$  \\
\cline{2-3}$$ & $100111000111$ &  $00011001001$  \\
\cline{2-3}$$ & $100111011001$ &  $01001000011$  \\
\cline{2-3}$$ & $100111100101$ &  $00011010001$  \\
\cline{2-3}$$ & $100111110111$ &  $00001001101$  \\
\cline{2-3}$$ & $101000000001$ &  $11111101111$  \\
\cline{2-3}$$ & $101000000111$ &  $10001111001$  \\
\cline{2-3}$$ & $101000010011$ &  $01111010010$  \\
\cline{2-3}$$ & $101000010101$ &  $00010000100$  \\
\cline{2-3}$$ & $101000101001$ &  $10100110101$  \\
\cline{2-3}$$ & $101001001001$ &  $10001100111$  \\
\cline{2-3}$$ & $101001100001$ &  $00101101110$  \\
\hline
\end{tabular}

\begin{tabular}{|c|c|c|}
\hline
\# cells & Primitive Poly. & CA-rule vector      \\
\hline $11$ & $101001101101$ &  $00110101110$  \\
\cline{2-3}$$ & $101001111001$ &  $01011011100$  \\
\cline{2-3}$$ & $101001111111$ &  $10000111011$  \\
\cline{2-3}$$ & $101010000101$ &  $00000011000$  \\
\cline{2-3}$$ & $101010010001$ &  $01011010110$  \\
\cline{2-3}$$ & $101010011101$ &  $00000001100$  \\
\cline{2-3}$$ & $101010100111$ &  $10100011101$  \\
\cline{2-3}$$ & $101010101011$ &  $01010111010$  \\
\cline{2-3}$$ & $101010110011$ &  $10111010001$  \\
\cline{2-3}$$ & $101010110101$ &  $10011001101$  \\
\cline{2-3}$$ & $101011010101$ &  $11011001001$  \\
\cline{2-3}$$ & $101011011111$ &  $11010011001$  \\
\cline{2-3}$$ & $101011101001$ &  $11010001101$  \\
\cline{2-3}$$ & $101011101111$ &  $10110100011$  \\
\cline{2-3}$$ & $101011110001$ &  $11010110001$  \\
\cline{2-3}$$ & $101011111011$ &  $00000000110$  \\
\cline{2-3}$$ & $101100000011$ &  $01010010111$  \\
\cline{2-3}$$ & $101100001001$ &  $00110101101$  \\
\cline{2-3}$$ & $101100010001$ &  $01111111111$  \\
\cline{2-3}$$ & $101100110011$ &  $00001111101$  \\
\cline{2-3}$$ & $101100111111$ &  $01011010011$  \\
\cline{2-3}$$ & $101101000001$  & $01100101101$ \\
\cline{2-3} $$ & $101101001011$ &  $00001011111$  \\
\cline{2-3} $$ & $101101011001$ &  $01011001101$  \\
\cline{2-3} $$ & $101101011111$ &  $00101010111$  \\
\cline{2-3} $$ & $101101100101$ &  $10100101110$  \\
\cline{2-3} $$ & $101101101111$ &  $00101101011$  \\
\cline{2-3} $$ & $101101111101$ &  $01010011101$  \\
\cline{2-3} $$ & $101110000111$ &  $00011101101$  \\
\cline{2-3} $$ & $101110001011$ &  $00111101001$  \\
\cline{2-3} $$ & $101110010011$ &  $00000100001$  \\
\cline{2-3} $$ & $101110010101$ &  $00010000001$  \\
\cline{2-3} $$ & $101110101111$ &  $01010101011$  \\
\cline{2-3} $$ & $101110110111$ &  $11000101110$  \\
\cline{2-3} $$ & $101110111101$ &  $01000110111$  \\
\hline
\end{tabular}
\quad
\begin{tabular}{|c|c|c|}
\hline
\# cells & Primitive Poly. & CA-rule vector\\
\hline $11$ & $101111001001$ &  $00110100111$  \\
\cline{2-3} $$ & $101111011011$ &  $00100111011$  \\
\cline{2-3} $$ & $101111011101$ &  $00010101111$  \\
\cline{2-3} $$ & $101111100111$ &  $00101001111$  \\
\cline{2-3} $$ & $101111101101$ &  $00100101111$  \\
\cline{2-3} $$ & $110000001011$ &  $01100001110$  \\
\cline{2-3} $$ & $110000001101$ &  $00000010000$  \\
\cline{2-3} $$ & $110000011001$ &  $11011110111$  \\
\cline{2-3} $$ & $110000011111$ &  $10000101101$  \\
\cline{2-3} $$ & $110001010111$ &  $00011001110$  \\
\cline{2-3} $$ & $110001100001$ &  $10001010011$  \\
\cline{2-3} $$ & $110001101011$ &  $10001001011$  \\
\cline{2-3} $$ & $110001110011$ &  $00100011110$  \\
\cline{2-3} $$ & $110010000101$ &  $00101001110$  \\
\cline{2-3} $$ & $110010001001$ &  $10011111111$  \\
\cline{2-3} $$ & $110010010111$ &  $00011010110$  \\
\cline{2-3} $$ & $110010011011$ &  $00101110010$  \\
\cline{2-3} $$ & $110010011101$ &  $00111010010$  \\
\cline{2-3} $$ & $110010110011$ &  $00100110110$  \\
\cline{2-3} $$ & $110010111111$ &  $00101100110$  \\
\cline{2-3} $$ & $110011000111$ &  $01000111010$  \\
\cline{2-3} $$& $110011001101$ &  $10100010101$      \\
\cline{2-3}$$ & $110011010011$ &  $01000101110$      \\
\cline{2-3}$$ & $110011010101$ &  $01001100110$      \\
\cline{2-3}$$ & $110011100011$ &  $10100101001$      \\
\cline{2-3}$$ & $110011101001$ &  $11001111111$      \\
\cline{2-3}$$ & $110011110111$ &  $11100111111$      \\
\cline{2-3}$$ & $110100000011$ &  $01001010011$      \\
\cline{2-3}$$ & $110100001111$ &  $00010110101$      \\
\cline{2-3}$$ & $110100011101$ &  $00000000001$      \\
\cline{2-3}$$ & $110100100111$ &  $00110100101$      \\
\cline{2-3}$$ & $110100101101$ &  $11110000010$      \\
\cline{2-3}$$ & $110101000001$ &  $00010101011$      \\
\cline{2-3}$$ & $110101000111$ &  $01001100101$      \\
\cline{2-3}$$ & $110101010101$ &  $00000101111$      \\
\hline
\end{tabular}

\begin{tabular}{|c|c|c|}
\hline
\# cells & Primitive Poly. & CA-rule vector      \\
\hline $11$ & $110101011001$ &  $01001011001$      \\
\cline{2-3}$$ & $110101100011$ &  $00101001011$      \\
\cline{2-3}$$ & $110101101111$ &  $01011010001$      \\
\cline{2-3}$$ & $110101110001$ &  $10101100010$      \\
\cline{2-3}$$ & $110110010011$ &  $00001101011$      \\
\cline{2-3}$$ & $110110011111$ &  $00001011011$      \\
\cline{2-3}$$ & $110110101001$ &  $00011010011$      \\
\cline{2-3}$$ & $110110111011$ &  $00100011011$      \\
\cline{2-3}$$ & $110110111101$ &  $00001010111$      \\
\cline{2-3}$$ & $110111001001$ &  $00101010101$      \\
\cline{2-3}$$ & $110111010111$ &  $00111010001$      \\
\cline{2-3}$$ & $110111011011$ &  $00010011101$      \\
\cline{2-3}$$ & $110111100001$ &  $01111011111$      \\
\cline{2-3}$$ & $110111100111$ &  $01111110111$      \\
\cline{2-3}$$ & $110111110101$ &  $11100010010$      \\
\cline{2-3}$$ & $111000000101$ &  $10000100010$      \\
\cline{2-3}$$ & $111000011101$ &  $00110011111$      \\
\cline{2-3}$$ & $111000100001$ &  $01011101101$      \\
\cline{2-3}$$ & $111000100111$ &  $00011111011$      \\
\cline{2-3}$$ & $111000101011$ &  $01111110001$      \\
\cline{2-3}$$ & $111000110011$ &  $00011011111$      \\
\cline{2-3}$$ & $111000111001$ & $01101011101$        \\
\cline{2-3} $$ & $111001000111$ & $11110001110$        \\
\cline{2-3} $$ & $111001001011$ & $00011111101$        \\
\cline{2-3} $$ & $111001010101$ & $00010001001$        \\
\cline{2-3} $$ & $111001011111$ & $00001001001$        \\
\cline{2-3} $$ & $111001110001$ & $01101010111$        \\
\cline{2-3} $$ & $111001111011$ & $10011111100$        \\
\cline{2-3} $$ & $111001111101$ & $00001000101$        \\
\cline{2-3} $$ & $111010000001$ & $01001101111$        \\
\cline{2-3} $$ & $111010010011$ & $01110100111$        \\

\hline
\end{tabular}
\quad
\begin{tabular}{|c|c|c|}
\hline
\# cells & Primitive Poly. & CA-rule vector\\
\hline    $11$ & $111011011101$ & $01111001101$        \\
\cline{2-3} $$ & $111010011111$ & $01100101111$        \\
\cline{2-3} $$ & $111010100011$ & $01101100111$        \\
\cline{2-3} $$ & $111010111011$ & $00111101101$        \\
\cline{2-3} $$ & $111011001111$ & $01110111001$        \\
\cline{2-3} $$ & $111011110011$ & $00101110111$        \\
\cline{2-3} $$ & $111011111001$ & $01111101001$        \\
\cline{2-3} $$ & $111100001011$ & $10111100101$        \\
\cline{2-3} $$ & $111100011001$ & $00001010100$        \\
\cline{2-3} $$ & $111100110001$ & $10110011101$        \\
\cline{2-3} $$ & $111100110111$ & $10000001001$        \\
\cline{2-3} $$ & $111101011101$ & $10110110011$        \\
\cline{2-3} $$ & $111101101011$ & $00111101110$        \\
\cline{2-3} $$ & $111101101101$ & $11001011101$        \\
\cline{2-3} $$ & $111101110101$ & $10010111011$        \\
\cline{2-3} $$ & $111110000011$ & $00010000110$        \\
\cline{2-3} $$ & $111110010001$ & $10110101011$        \\
\cline{2-3} $$ & $111110010111$ & $10101011011$        \\
\cline{2-3} $$ & $111110011011$ & $11110011001$        \\
\cline{2-3} $$ & $111110100111$ & $10001110111$        \\
\cline{2-3} $$ & $111110101101$ & $10001101111$        \\
\cline{2-3} $$ & $111110110101$ & $10110001111$        \\
\cline{2-3} $$ & $111111001101$ & $00010011000$        \\
\cline{2-3} $$ & $111111010011$ & $11011000111$        \\
\cline{2-3} $$ & $111111100101$ & $11001100111$        \\
\cline{2-3} $$ & $111111101001$ & $01011110110$        \\
\hline $12$ & $1111110011001$ & $010010110010$\\
\cline{2-3} $$ & $1111110111011$ & $100000110101$\\
\cline{2-3} $$ & $1111110111101$ & $010010010110$\\
\cline{2-3} $$ & $1111111001001$ & $111000000011$\\
\hline
\end{tabular}

\begin{tabular}{|c|c|c|}
\hline
\# cells & Primitive Poly. & CA-rule vector\\
\hline  $12$   & $1000001010011$ & $011011000110$\\
\cline{2-3} $$ & $1000001101001$ & $100101100101$\\
\cline{2-3} $$ & $1000001111011$ & $011100110010$\\
\cline{2-3} $$ & $1000001111101$ & $000001000100$\\
\cline{2-3} $$ & $1000010011001$ & $100101010011$\\
\cline{2-3} $$ & $1000011010001$ & $001100111010$\\
\cline{2-3} $$ & $1000011101011$ & $001110100110$\\
\cline{2-3} $$ & $1000100000111$ & $101001010101$\\
\cline{2-3} $$ & $1000100011111$ & $110000101011$\\
\cline{2-3} $$ & $1000100100011$ & $100100011101$\\
\cline{2-3} $$ & $1000100111011$ & $000111101100$\\
\cline{2-3} $$ & $1000101001111$ & $111101110111$\\
\cline{2-3} $$ & $1000101010111$ & $100111001001$\\
\cline{2-3} $$ & $1000101100001$ & $111011111011$\\
\cline{2-3} $$ & $1000101101011$ & $101100110001$\\
\cline{2-3} $$ & $1000110000101$ & $110001011001$\\
\cline{2-3} $$ & $1000110110011$ & $101100100011$\\
\cline{2-3} $$ & $1000111011001$ & $000011101110$\\
\cline{2-3} $$ & $1000111011111$ & $010101101100$\\
\cline{2-3} $$ & $1001000001101$ & $000100011111$\\
\cline{2-3} $$ & $1001000110111$ & $010101010101$\\
\cline{2-3} $$ & $1001000111101$ & $001000011111$\\
\cline{2-3} $$ & $1001001100111$ & $011100101001$\\
\cline{2-3} $$ & $1001001110011$ & $011010110001$\\
\cline{2-3} $$ & $1001001111111$ & $000011110011$\\
\cline{2-3} $$ & $1001010111001$ & $001101001101$\\
\cline{2-3} $$ & $1001011000001$ & $010011010011$\\
\cline{2-3} $$ & $1001011001011$ & $000101101101$\\
\cline{2-3} $$ & $1001100001111$ & $001101000111$\\
\cline{2-3} $$ & $1001100011101$ & $000101110011$\\
\cline{2-3} $$ & $1001100100001$ & $110111000100$\\
\cline{2-3} $$ & $1001100111001$ & $001011100011$\\
\cline{2-3} $$ & $1001100111111$ & $110100011100$\\
\cline{2-3} $$ & $1001101001101$ & $010010101101$\\
\cline{2-3} $$ & $1001101110001$ & $100101101010$\\
\hline
\end{tabular}
\quad
\begin{tabular}{|c|c|c|}
\hline
\# cells & Primitive Poly. & CA-rule vector\\
\hline     $12$ & $1001110011001$ & $001000111101$\\
\cline{2-3} $$ & $1001110100011$ & $000011011101$\\
\cline{2-3} $$ & $1001110101001$ & $100010111100$\\
\cline{2-3} $$ & $1010000000111$ & $010001010010$\\
\cline{2-3} $$ & $1010000110001$ & $000011100100$\\
\cline{2-3} $$ & $1010000110111$ & $111100011011$\\
\cline{2-3} $$ & $1010001001111$ & $111101000111$\\
\cline{2-3} $$ & $1010001011101$ & $101111010101$\\
\cline{2-3} $$ & $1010001100111$ & $111110001011$\\
\cline{2-3} $$ & $1010001110101$ & $000001110100$\\
\cline{2-3} $$ & $1010010100111$ & $001010100010$\\
\cline{2-3} $$ & $1010010101101$ & $001010001010$\\
\cline{2-3} $$ & $1010011010011$ & $110101110101$\\
\cline{2-3} $$ & $1010100001111$ & $011001111110$\\
\cline{2-3} $$ & $1010100011101$ & $100110000001$\\
\cline{2-3} $$ & $1010101001101$ & $110110110011$\\
\cline{2-3} $$ & $1010110010011$ & $000001101010$\\
\cline{2-3} $$ & $1010111000101$ & $001010000110$\\
\cline{2-3} $$ & $1010111010111$ & $000101100010$\\
\cline{2-3} $$ & $1010111011101$ & $101101110011$\\
\cline{2-3} $$ & $1010111101011$ & $101001011111$\\
\cline{2-3} $$ & $1011000001001$ & $010101011111$\\
\cline{2-3} $$ & $1011001000111$ & $111101011010$\\
\cline{2-3} $$ & $1011001010101$ & $000110001001$\\
\cline{2-3} $$ & $1011001011001$ & $111010111010$\\
\cline{2-3} $$ & $1011010100101$ & $000100100011$\\
\cline{2-3} $$ & $1011010111101$ & $011111110001$\\
\cline{2-3} $$ & $1011100010101$ & $011010011111$\\
\cline{2-3} $$ & $1011100011001$ & $000001010101$\\
\cline{2-3} $$ & $1011101000011$ & $001001001001$\\
\cline{2-3} $$ & $1011101000101$ & $000010100101$\\
\cline{2-3} $$ & $1011101110101$ & $011011110011$\\
\cline{2-3} $$ & $1011110001001$ & $010100000101$\\
\cline{2-3} $$ & $1011110101101$ & $010111111001$\\
\cline{2-3} $$ & $1011110110011$ & $010010001001$\\
\hline
\end{tabular}

\begin{tabular}{|c|c|c|}
\hline
\# cells & Primitive Poly. & CA-rule vector\\
\hline      $12$ & $1011110111111$ & $010010010001$\\
\cline{2-3} $$ & $1011111000001$ & $011100000001$\\
\cline{2-3} $$ & $1100001010111$ & $100111001101$\\
\cline{2-3} $$ & $1100001011101$ & $111010010011$\\
\cline{2-3} $$ & $1100010010001$ & $000101000010$\\
\cline{2-3} $$ & $1100010010111$ & $000010001010$\\
\cline{2-3} $$ & $1100010111001$ & $001010000010$\\
\cline{2-3} $$ & $1100011101111$ & $100000111111$\\
\cline{2-3} $$ & $1100100011011$ & $111000111001$\\
\cline{2-3} $$ & $1100100110101$ & $001111101010$\\
\cline{2-3} $$ & $1100101000001$ & $000000010110$\\
\cline{2-3} $$ & $1100101100101$ & $001110101110$\\
\cline{2-3} $$ & $1100101111011$ & $101011010011$\\
\cline{2-3} $$ & $1100110001011$ & $001000001100$\\
\cline{2-3} $$ & $1100110110001$ & $110010001111$\\
\cline{2-3} $$ & $1100110111101$ & $000111111100$\\
\cline{2-3} $$ & $1100111001001$ & $011111001010$\\
\cline{2-3} $$ & $1100111001111$ & $101101101001$\\
\cline{2-3} $$ & $1100111100111$ & $110101001011$\\
\cline{2-3} $$ & $1101000011011$ & $000100000101$\\
\cline{2-3} $$ & $1101000101011$ & $010010101111$\\
\cline{2-3} $$ & $1101000110011$ & $000101000001$\\
\cline{2-3} $$ & $1101001101001$ & $010111001011$\\
\cline{2-3} $$ & $1101010001011$ & $101011110010$\\
\cline{2-3} $$ & $1101011010001$ & $001101011011$\\
\cline{2-3} $$ & $1101011100001$ & $001111000111$\\
\cline{2-3} $$ & $1101011110101$ & $010011011101$\\
\cline{2-3} $$ & $1101100001011$ & $001011110011$\\
\cline{2-3} $$ & $1101100010011$ & $011100011101$\\
\cline{2-3} $$ & $1101100011111$ & $000001111111$\\
\cline{2-3} $$ & $1101101010111$ & $011100111001$\\
\cline{2-3} $$ & $1101110010001$ & $010010011111$\\
\cline{2-3} $$ & $1101110100111$ & $011010001111$\\
\cline{2-3} $$ & $1101110111111$ & $011100010111$\\
\cline{2-3} $$ & $1101111000001$ & $000110111101$\\

\hline
\end{tabular}
\quad
\begin{tabular}{|c|c|c|}
\hline
\# cells & Primitive Poly. & CA-rule vector\\
\hline     $12$ & $1101111010011$ & $010110101011$\\
\cline{2-3} $$ & $1110000000101$ & $000000011111$\\
\cline{2-3} $$ & $1110000010001$ & $001100001011$\\
\cline{2-3} $$ & $1110000010111$ & $001111111011$\\
\cline{2-3} $$ & $1110000100111$ & $010101000101$\\
\cline{2-3} $$ & $1110001001101$ & $010100100101$\\
\cline{2-3} $$ & $1110010000111$ & $001110001001$\\
\cline{2-3} $$ & $1110010011111$ & $000001011101$\\
\cline{2-3} $$ & $1110010100101$ & $001010100101$\\
\cline{2-3} $$ & $1110010111011$ & $000011011001$\\
\cline{2-3} $$ & $1110011000101$ & $000110011001$\\
\cline{2-3} $$ & $1110011001001$ & $000101110001$\\
\cline{2-3} $$ & $1110011001111$ & $011001000011$\\
\cline{2-3} $$ & $1110011110011$ & $000001101101$\\
\cline{2-3} $$ & $1110100000111$ & $011111101011$\\
\cline{2-3} $$ & $1110100100011$ & $000010101101$\\
\cline{2-3} $$ & $1110101000011$ & $000101001101$\\
\cline{2-3} $$ & $1110101010001$ & $001010110001$\\
\cline{2-3} $$ & $1110101011011$ & $001101010001$\\
\cline{2-3} $$ & $1110101110101$ & $001011000101$\\
\cline{2-3} $$ & $1110110000101$ & $000111000011$\\
\cline{2-3} $$ & $1110110001001$ & $000011100011$\\
\cline{2-3} $$ & $1111000010101$ & $011011111110$\\
\cline{2-3} $$ & $1111000011001$ & $111010110111$\\
\cline{2-3} $$ & $1111000101111$ & $011100010010$\\
\cline{2-3} $$ & $1111001000101$ & $010111111110$\\
\cline{2-3} $$ & $1111001010001$ & $000101101100$\\
\cline{2-3} $$ & $1111001100111$ & $100100101001$\\
\cline{2-3} $$ & $1111001110011$ & $110000001011$\\
\cline{2-3} $$ & $1111010001111$ & $000000111110$\\
\cline{2-3} $$ & $1111011100011$ & $001000111010$\\
\cline{2-3} $$ & $1111100010001$ & $110010001001$\\
\cline{2-3} $$ & $1111100011011$ & $000101010110$\\
\cline{2-3} $$ & $1111100100111$ & $001100001110$\\
\cline{2-3} $$ & $1111101110001$ & $001001011010$\\
\hline
\end{tabular}

\textbf{Conclusion} In this paper a simple algorithm to compute
rule vectors for $n$-cell maximum length CA has been introduced.
Also, all maximum length CA rule vectors for cell size $2$ to $12$
have been computed by employing proposed algorithm and they have
been tabulated. Programmable rule vectors based maximum length CA
can be used to design cryptographic primitives. Since the list of
all rule vectors are available so it will certainly reduce design
cycle time.

\end{document}